\documentclass[aps,prl, twocolumn,showpacs,superscriptaddress]{revtex4-1}  % for double-spaced preprint
\usepackage{graphicx}  % needed for figures
\usepackage{subfigure}
\usepackage{dcolumn}   % needed for some tables
\usepackage{bm}        % for math
\usepackage{amssymb,amsmath,amsopn,amsfonts}
\usepackage{colordvi}
\usepackage{color}
\usepackage[none]{hyphenat}
\usepackage{overpic}
\usepackage{amsmath}
\usepackage{amsthm}
\usepackage{amssymb}
\usepackage{bm}
\usepackage{bbm}

\newtheorem{lemma}{Lemma}

\newcommand{\ket}[1]{\ensuremath{\left| #1 \right\rangle}}
\newcommand{\bra}[1]{\ensuremath{\left\langle #1 \right|}}

\newcommand{\ketbra}[2]{\ensuremath{\left| #1 \rangle \langle #2\right|}}

\begin{document}
\widetext
\title{Experimental cyclic inter-conversion between Coherence and Quantum Correlations}
\date{\today}

\author{Kang-Da Wu}
\affiliation{CAS Key Laboratory of Quantum Information, University of Science and Technology of China, Hefei, 230026, People's Republic of China}
\affiliation{Synergetic Innovation Center of Quantum Information and Quantum Physics}

\author{Zhibo Hou}
\affiliation{CAS Key Laboratory of Quantum Information, University of Science and Technology of China, Hefei, 230026, People's Republic of China}
\affiliation{Synergetic Innovation Center of Quantum Information and Quantum Physics}

\author{Yuan-Yuan Zhao}
\affiliation{CAS Key Laboratory of Quantum Information, University of Science and Technology of China, Hefei, 230026, People's Republic of China}
\affiliation{Synergetic Innovation Center of Quantum Information and Quantum Physics}

\author{Guo-Yong Xiang}
\email{gyxiang@ustc.edu.cn}
\affiliation{CAS Key Laboratory of Quantum Information, University of Science and Technology of China, Hefei, 230026, People's Republic of China}
\affiliation{Synergetic Innovation Center of Quantum Information and Quantum Physics}

\author{Chuan-Feng Li}
\affiliation{CAS Key Laboratory of Quantum Information, University of Science and Technology of China, Hefei, 230026, People's Republic of China}
\affiliation{Synergetic Innovation Center of Quantum Information and Quantum Physics}

\author{Guang-Can Guo}
\affiliation{CAS Key Laboratory of Quantum Information, University of Science and Technology of China, Hefei, 230026, People's Republic of China}
\affiliation{Synergetic Innovation Center of Quantum Information and Quantum Physics}

\author{Jiajun Ma}
\affiliation{Center for Quantum Information, Institute for Interdisciplinary Information Sciences, Tsinghua University, Beijing, China}

\author{Qiong-Yi He}
\affiliation{ State Key Laboratory of Mesoscopic Physics, School of Physics, Peking University, Collaborative Innovation Center of Quantum Matter, Beijing 100871, China}

\author{Jayne Thompson}
\affiliation{Centre for Quantum Technologies, National University of Singapore, Singapore}

\author{Mile Gu}
\email{gumile@ntu.edu.sg}
\affiliation{School of Mathematical and Physical Sciences, Nanyang Technological University, Singapore}
\affiliation{Complexity Institute, Nanyang Technological University, Singapore}
\affiliation{Centre for Quantum Technologies, National University of Singapore, Singapore}

\begin{abstract}
Quantum resource theories seek to quantify sources of non-classicality that bestow quantum technologies their operational advantage. Chief among these are studies of quantum correlations and quantum coherence. The former to isolate non-classicality in the correlations between systems, the latter to capture non-classicality of quantum superpositions within a single physical system. Here we present a scheme that cyclically inter-converts between these resources without loss. The first stage converts coherence present in an input system into correlations with an ancilla. The second stage harnesses these correlations to restore coherence on the input system by measurement of the ancilla. We experimentally demonstrate this inter-conversion process using linear optics. Our experiment highlights the connection between non-classicality of correlations and non-classicality within local quantum systems, and provides potential flexibilities in exploiting one resource to perform tasks normally associated with the other.
\end{abstract}

%\pacs{03.65.Wj, 02.50.-r, 03.67.-a}
%\pacs{03.65.Wj, 06.20.Dk, 42.25.Ja, 03.67.-a}
\maketitle

%For nearly half a century, quantum mechanics is not only going deep into many branches of physics but also in many other fields such as chemical, biological and information science. The result of this in-depth application of quantum mechanics is that it has greatly promoted the development of these subjects and the formation of many hot spots of scientific research. One of the hot spots is quantum information science, which lies in the heart of quantum computation and quantum communication.

%The discovery of physical processes that convert one resource into another has led to significant scientific progress. The demonstration that kinetic energy can be converted to thermal energy, for example, led to rapid developments in thermodynamics, and elevated conservation of energy to the role of a fundamental physical principle \cite{mach2014history}. Understanding such inter-conversions allows us to validate potential new resources in the context of more established resource theories, and inspires fresh perspectives on how such resources lead to practical benefit.

In quantum science, there is continuing interest in isolating resources that quantify features of non-classicality, and thus underly advantaged quantum information processing. Entanglement is an iconic example, representing correlations in multipartite systems with no classical counterpart. The subsequent discovery that quantum advantage can persist in entanglement-breaking regimes propelled interest in more general forms of non-classicality. Quantum discord became a prominent candidate, quantifying quantum correlations that may persist in unentangled systems~\cite{HendV01,ollivier2001quantum,modi2012classical}. Resource theoretic perspectives also motivated revisiting long-standing notions of non-classicality in localized systems, where quantum superposition states such as Schrodinger's cat stand as key examples. This resulted in the resource theory of quantum coherence, where one identifies a natural classical basis (e.g. alive and dead states of the cat), and a state is said to contain coherence if it cannot be expressed as a mixture of such classical states~\cite{QuantifyingCoherence,streltsov2016quantum,fancoherence2017}. These studies proved invaluable for isolating how non-classical resources beyond entanglement can potentially empower quantum technologies~\cite{weedbrook2016discord,lanyon2008experimental,gu2012observing,yuan2015intrinsic,karlstrom2011increasing,levi2014quantitative,OperationalResourceTheoryofCoherence}.
%While most studies treat non-classicality of correlations and coherence of singular systems separately,

Recently there has been growing evidence that in certain settings these resources can be interconverted, allowing us to consume local non-classicality to create quantum correlations with an ancillary system and vice versa~\cite{killoran2016converting,MeasuringQuantumCoherencewithEntanglement,ConvertingCoherencetoQuantumCorrelations,AssistedDistillationofQuantumCoherence,fanhuscireport2017}. These studies field dual benefits. They can provide ways of validating potential new resources in the context of more established resource theories, and inspire fresh perspectives on how such resources can lead to practical advantage.

Here we leverage these ideas by designing a quantum circuit that cyclically inter-converts between coherence and discord and implement it using discrete photonics. The first half cycle experimentally realizes a recently proposed protocol for converting coherence to discord~\cite{ConvertingCoherencetoQuantumCorrelations}, where coherence on one input system is converted to discordant correlations between it and a suitable ancilla. The second adapts theoreticical and experimental techniques in coherence distillation and steering~\cite{AssistedDistillationofQuantumCoherence,fanhuscireport2017,wu2017experimentally}, whereby we restore coherence in the original input system via measurement of the correlated ancilla. In principle, this cycle can be completed without any loss of resource, and operates on any input state with non-zero coherence. Experimentally, we implement this for the case where the input and ancilla are single qubits, and demonstrate up to 80 percent of initial coherence is retained within a single cycle. Our work illustrates that under appropriate conditions, coherence and discord behave analogously to different forms of energy, exchangeable from one form to another.

%There exist multiple settings in which coherence within a local system can be converted to non-classical correlations, either in the form of entanglement or discord~\cite{killoran2016converting,MeasuringQuantumCoherencewithEntanglement,ConvertingCoherencetoQuantumCorrelations}. Meanwhile, recent studies show that when local operations that generate coherence on a system are forbidden, we can still induce non-classicality by leveraging correlations~\cite{AlexanderDistributedCoherence,AssistedDistillationofQuantumCoherence,fanhuscireport2017,wu2017experimentally}. One can steer the system towards more non-classical states by measuring a suitable ancilla, provided that ancilla is correlated with the system of interest.

\begin{figure}[htp]
\includegraphics[scale=0.34]{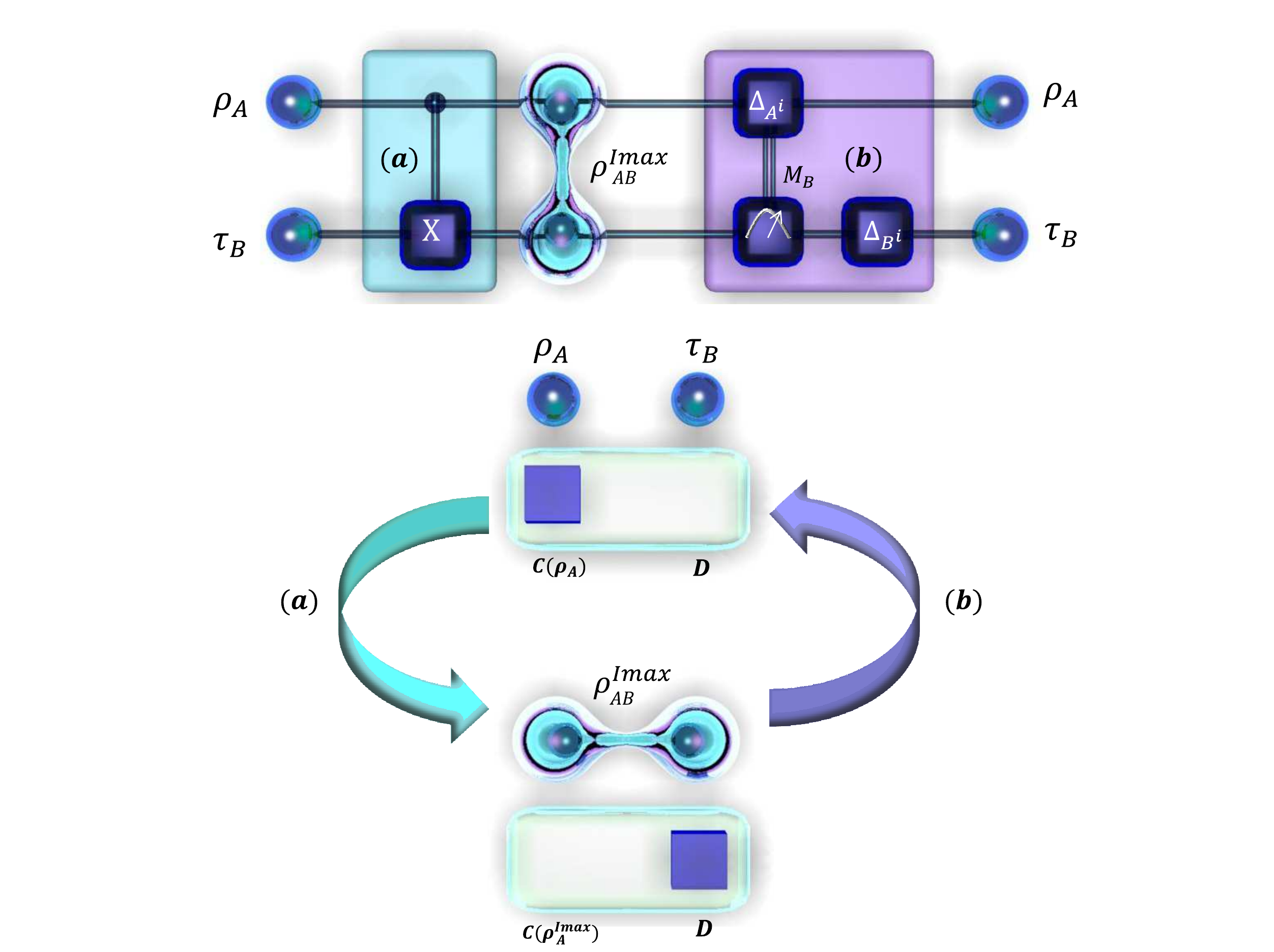}
\caption{\label{fig:theoretical} \textbf{Cyclic inter-conversion between coherence and discord.} The cycle can be divided into two stages. In (a) coherence in system $A$ is converted into discord between A and B using incoherent operations. In (b) the coherence in $A$ is restored by application of LQICC operations on the output from stage (a).}\label{fig:theoreticalscheme}
\end{figure}

{\bf Background.}
We first review coherence and discord. The resource theory of coherence formalizes the intuition that quantum superpositions are non-classical~\cite{QuantifyingCoherence}. In this framework, some orthogonal basis of states, referred to as the reference basis, $\{\ket{i}\}$ are considered classical, usually motivated by physical grounds of being easy to synthesize or store. For qubits, this basis is usually denoted \{$\ket{0},\ket{1}\}$ by convention, as they are often considered analogous to the two states of a classical bit. Any mixture of such states, $\chi=\sum_{i}p_{i}\ketbra{i}{i}$ with $\sum_{i}p_{i}=1$ and $p_{i}\geq0$, is considered classical and termed \emph{incoherent}. For bipartite systems partitioned by $A$ and $B$, each with respective reference basis $\{\ket{i}\}$ and $\{\ket{j}\}$, the incoherent states take the form $\chi_{AB}=\sum_{ij}p_{ij}\ketbra{i}{i}\otimes\ketbra{j}{j}$ with $\sum_{ij}p_{ij}=1$ and $p_{ij}\geq0$. The rationale being that such states have a 1-1 map to classical probability distributions on classical states. States lying outside this set contain coherence.

Incoherent operations define the set of operations that take incoherent states to incoherent states, representing `free operations' that cannot create non-classicality and thus can be simulated by classical stochastic maps~\cite{QuantifyingCoherence}. Each measure of coherence is then a real valued function on quantum states that cannot increase under incoherent operations~\cite{QuantifyingCoherence,girolami2014observable,yuan2015intrinsic}. Here, we adopt the relative entropy of coherence $C(\rho)=\min_{\chi\in\mathcal{I}}S(\rho||\chi)$, representing the relative entropy of $\rho$ to the closest incoherent state.

Quantum discord captures quantum correlations beyond entanglement in bipartite systems~\cite{ollivier2001quantum,henderson2001classical,modi2012classical}, and represents unavoidable disturbance induced by local measurement. For a bipartite system partitioned by $A$ and $B$, a state without discord takes the form $\sigma_{AB}=\sum_ip_{ij}\ketbra{i}{i}_A\otimes\ketbra{j}{j}_B$, where $\{p_{ij}\}\ge 0$, $\sum_{ij}p_{ij}=1$ and $\{\ket{i}_A\}$, $\{\ket{j}_B\}$ are any orthogonal basis for subsystems $A$ and $B$, respectively. Such states are referred to as being classically correlated (CC), and any other bipartite state is defined to have discord. There are several measures of discord in literature, among which we adopt the \emph{relative entropy of discord} defined as  $D(\rho_{AB})=\min_{\sigma_{AB}}S(\rho_{AB}||\sigma_{AB})$, where the minimum is taken over classically correlated $\sigma_{AB}$~\cite{modi2012classical}.

{\bf Framework.}
Our scheme for cyclic conversion between coherence and discord is illustrated in Fig. \ref{fig:theoretical}. We begin with a system, denoted $A$ in some arbitrary state $\rho_A$, with coherence $C(\rho_{A})$. We then introduce a suitable ancilla $B$. The cycle is then divided into two halves. The first half consumes the initial coherence in $A$ to generate discord between $A$ and $B$; the second destroys these correlations by measuring $B$, and uses the measurement results to steer $A$ into a state with coherence.

The first half-cycle adopts a recent proposal for converting coherence to discord. Using system $A$ as the only source of coherence, the proposal considers how much quantum correlations can be generated between $A$ and some initially incoherent ancilla $B$ using only incoherent operations. The amount of coherence initially possessed by $A$ then bounds the discord that can be induced between input and ancilla~\cite{ConvertingCoherencetoQuantumCorrelations}. Formally, let $\Delta_{AB}$ be an arbitrary incoherent operation, and $\rho_{AB}^I=\Delta_{AB}(\rho_{A}\otimes\tau_{B})$, then
\begin{equation}\label{discordbound}
D(\rho_{AB}^{I})\leq C(\rho_{A}),
\end{equation}
Our scheme employs a circuit where this conversion is maximally efficient, such that all coherence on $A$ is consumed and transformed into discord. The resulting transformed state $\rho_{AB}^{Imax}$ satisfies $D(\rho_{AB}^{Imax}) = C(\rho_{A})$ and $C(\rho^{Imax}_{A}) = 0$, where $\rho^{Imax}_A = {\rm tr}_B \rho^{Imax}_{AB}$. This is always possible for any input $\rho_A$, using appropriate choice of ancilla and the application of a generalized CNOT gate (See supplementary materials for details).

\begin{figure*}[htp]
\label{fig:experimentalsetup}
\includegraphics[scale=0.7]{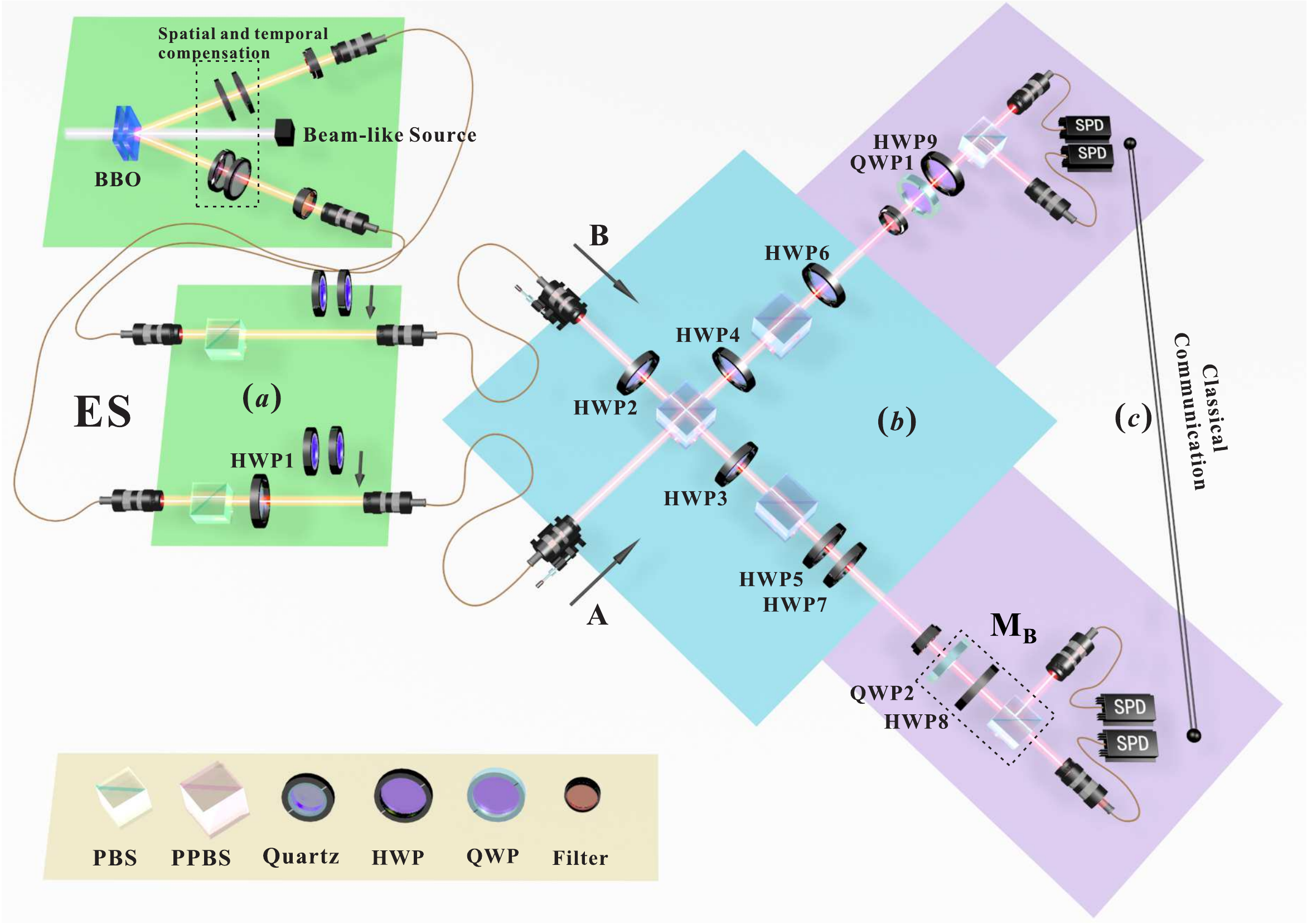}
\caption{\label{fig:experiment}\textbf{Experimental setup.} The setup divides into three modules, (a) state preparation, (b) coherence to discord conversion and (c) coherence restoration. Module (a) initializes one photon in some arbitrary state $\rho_{A}$ (named control), and a second photon in state $\ketbra{0}{0}_{B}$ (named target). Module (b) implements a CNOT gate between control and target, which converts the coherence in $\rho_{A}$ into discordant correlations between control and target photons. This results in discord state $\rho_{AB}^{Imax}$. Module (c) implements a von Nuemann measurement on the target photon by use of QWP2, HWP8 together with a PBS and two SPDs. The final coherence on $A$ is then detected by quantum state tomography using a combination of QWP1, HWP9, PBS and two SPDs. Note that at the end of module (b), we can also measure the discord in $\rho_{AB}^{Imax}$ via state tomography by using the aforementioned measurement apparatus. Key elements include: half-wave plate (HWP); quarter-wave plate (QWP); partially polarizing beam splitter (PPBS); polarizing beam splitter (PBS); single photon counting module (SPCM); entangled source (ES).}
%Rotation angles of HWPs: HWP2: $22.5^{\circ}$; HWP3: $45^{\circ}$; HWP4: $45^{\circ}$; HWP5: $45^{\circ}$; HWP6: $45^{\circ}$; HWP7: $22.5^{\circ}$. For detail see appendix.}
\end{figure*}

The second half-cycle uses discord in $\rho_{AB}^{Imax}$ to restore coherence in $A$ by suitable local measurements on $B$. To do so meaningfully, we forbid operations that create discord, or are able to spontaneously generate coherence in $A$. Specifically, we consider the class of operations -- known as local quantum incoherent operations and classical communication (LQICC) -- that involve measurement on $B$, and applying some incoherent operation to $A$ conditioned on the resulting outcome. Such a framework arose in the context of coherence distillation~\cite{AssistedDistillationofQuantumCoherence,fanhuscireport2017,AlexanderDistributedCoherence}, where it was motivated by the question of how correlations between $A$ and $B$ can be harnessed to steer $A$ into a more coherent state. For the specific case were the initial state is $\rho_{AB}^{Imax}$, we find
\begin{equation}\label{cohbound}
C(\rho_{A}^{\textmd{LQICC}})\leq D(\rho_{AB}^{Imax}) = C(\rho_{A})
\end{equation}
where $\rho_{A}^{\textmd{LQICC}}$ is the state of $A$ after any possible LQICC operation (See Supplementary Informations for proof). Thus the amount of coherence that can be restored to $A$ is upper bounded by the discord generated in the first half cycle. Meanwhile the measurement on $B$ destroys all discord between $A$ and $B$. In supplementary materials (section S1), we show this bound can be exactly saturated. Combining the two sub-protocols, we obtain a cyclic procedure that inter-converts between coherence and discord which can be repeated ad-infinitum.

{\bf Protocol.}
Our experiment implements such a cycle for the case where both $A$ and $B$ are qubits, and the basis $\{\ket{0},\ket{1}\}$ is considered classical. $A$ is initialized in state $\rho_A$ with coherence $C(\rho_A)$, while $B$ is initialized in state $\ket{0}$. The first stage converts the coherence in $A$ into discordant correlations between $A$ and $B$. This is achieved by a CNOT gate. The resulting state $\rho_{AB}^{Imax}$ has zero coherence on $A$ and discord $C(\rho_A)$, representing complete conversion. The second stage adopts techniques in coherence distillation to restore the coherence in $A$ by taking advantage of the discordant correlations created in the previous stage. This is achieved by measurement of $B$ mutually unbiased with Pauli $Z$ basis (actually we chose Pauli $Y$ basis in our experiment), with measurement outcomes $m \in \{0,1\}$. Knowledge of this outcome collapses system $A$ into a state with coherence $C(\rho_A)$ under ideal conditions.

%Conditioned on this outcome, application of operator $Z^m$, where $Z = \ket{0}\bra{0} - \ket{1}\bra{1}$ will restore all coherence to system $A$.
%or just dividing $A$'s photons for future arrangement according to the measurement outcomes allows us the restore a state on $A$ with coherence $C(\rho_A)$ under ideal theoretical conditions.

We realize this procedure using linear optics, where the  horizontal and vertical polarizations of a single photon denote the basis $\{\ket{0},\ket{1}\}$.  The experimental schematic is outlined in Fig. \ref{fig:experiment}, where we have divided the protocol into three self-contained modules. The first module (green) represents state preparation. The second module (cyan) converts coherence into discord, while the final module (purple) applies coherence distillation to restore initial coherence. For further details, see supplementary materials (section S2).

%For converting coherence to discord, we first implement CNOT gate as the physical realization of incoherent operation $\Delta_{AB}$. The Photons of A (control) and B (target) interference nonclassically at the first PPBS, together with the other two identical supplemental compensators, the gate is correctly implemented with probability $1/9$. After the implementation of incoherent gate operation $\Delta_{AB}$, the intermediate state $\rho_{AB}^{I}$ is directly analyzed (see two light purple rectangular areas in Fig. \ref{fig:experiment}). The tomographied data is used to reconstruct the quantum state $\rho_{AB}^{I}$ and calculate the discord  $D(\rho_{AB}^{I})$ for the verification of Eq. (\ref{discordbound}).
%
%As for the realization of recycling discord back to coherence on A, the LQICC operations need to be constructed. As discussed in Appendix \ref{sec:appendix1}, we can get the maximal increase in coherence on A's side via local measurement on B and one-way classical communication based on the broadcast of B. Thus, the lower light purple area in Fig. \ref{fig:experiment} consisting of wave plates and PBS are used to perform local measurement on B while on A's side, the upper one is used to perform quantum state tomography on A for detecting the final coherence $C(\rho_{A}^{\textmd{LQICC}})$. Actually, the optimal von Nuemann measurement on B ($M_{B}$) is chosen as $\sigma_{y}$ in our experiment.

\begin{figure}[htp]
\label{fig:result}
\includegraphics[scale=0.41]{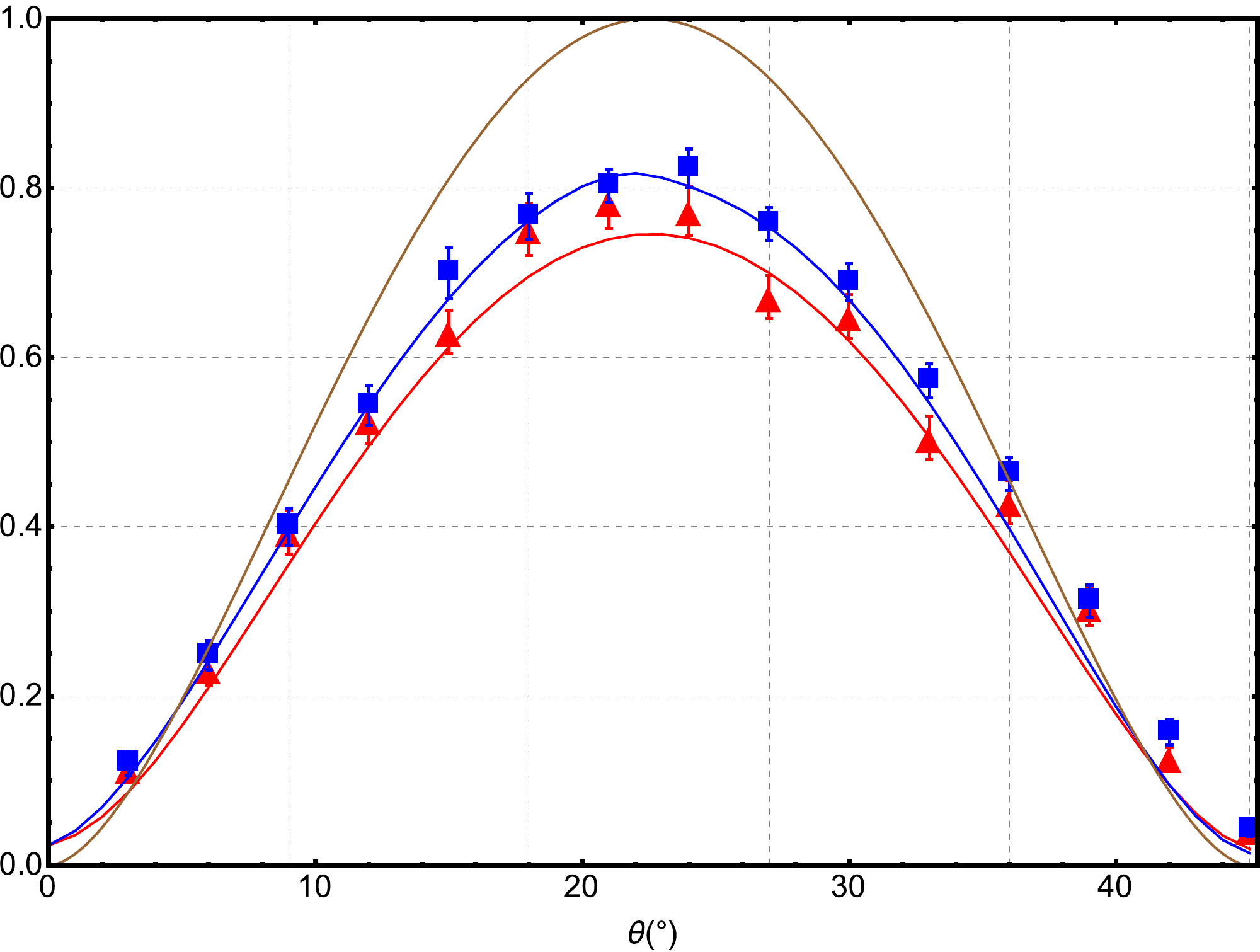}
\caption{\label{fig:purestate} \textbf{Discord generated and coherence restored for various pure state inputs.} We conducted the inter-conversion protocol for various pure state inputs of the form $\ket{\phi}_{AB}=(\cos2\theta\ket{0}_{A}+\sin2\theta\ket{1}_{A})\ket{0}_{B}$, for various values of $\theta$ between $0^\circ$ and $45^\circ$. The coherence of this input state is given by $h(\cos^{2}2\theta)$, where $h(p) = - p \log p - (1-p) \log (1-p)$ is the binary entropy (brown line). The first stage of the experiments converts this coherent into discord (blue squares). After the second stage of the experimental, the amount of coherence restored in the input photon is given by the red triangles. These experimental results agree well theory, once the imperfect fidelity of our CNOT gate (of approximately $0.885$) is taken into account (blue and red lines).}
\end{figure}

{\bf Results.}
We test the cyclic inter-conversion for both pure and mixed inputs $\rho_A$. Fig. \ref{fig:purestate} illustrates the results when $A$ is initialized in some pure state $\ket{\phi}_{A}=\cos2\theta\ket{0}+\sin2\theta\ket{1}$, with initial coherence as depicted by the solid brown line. After application of CNOT, the discord of the resulting state $\rho_{AB}^{Imax}$ can be retrieved by density matrix reconstruction (see Supplementary Informations for details). We see that the vast majority of this coherence is converted into discord (blue squares). The next stage of the experiment restores the coherence of $A$ by measurement of $B$. This involves a variant of experimental techniques developed for coherence distillation~\cite{wu2017experimentally}, where we retrieve most of the coherence originally in $\rho_A$ (see red triangles).

Note that under ideal scenarios, this conversion process would be lossless. In our experiment, we experienced a loss of approximately 20 percent. To understand this divergence, we performed quantum process tomography (QPT) \cite{Obri04quantum} to ascertain the process fidelity of the CNOT gate used during the conversion protocol (see Supplementary Informations for details).  Once the achieved fidelity of $0.885$ is taken into account, theoretical predictions and experimental results agree to within reasonable accuracy (see red and blue lines in Fig. \ref{fig:purestate}).

\begin{figure}[htp]
\label{fig:result}
\includegraphics[scale=0.41]{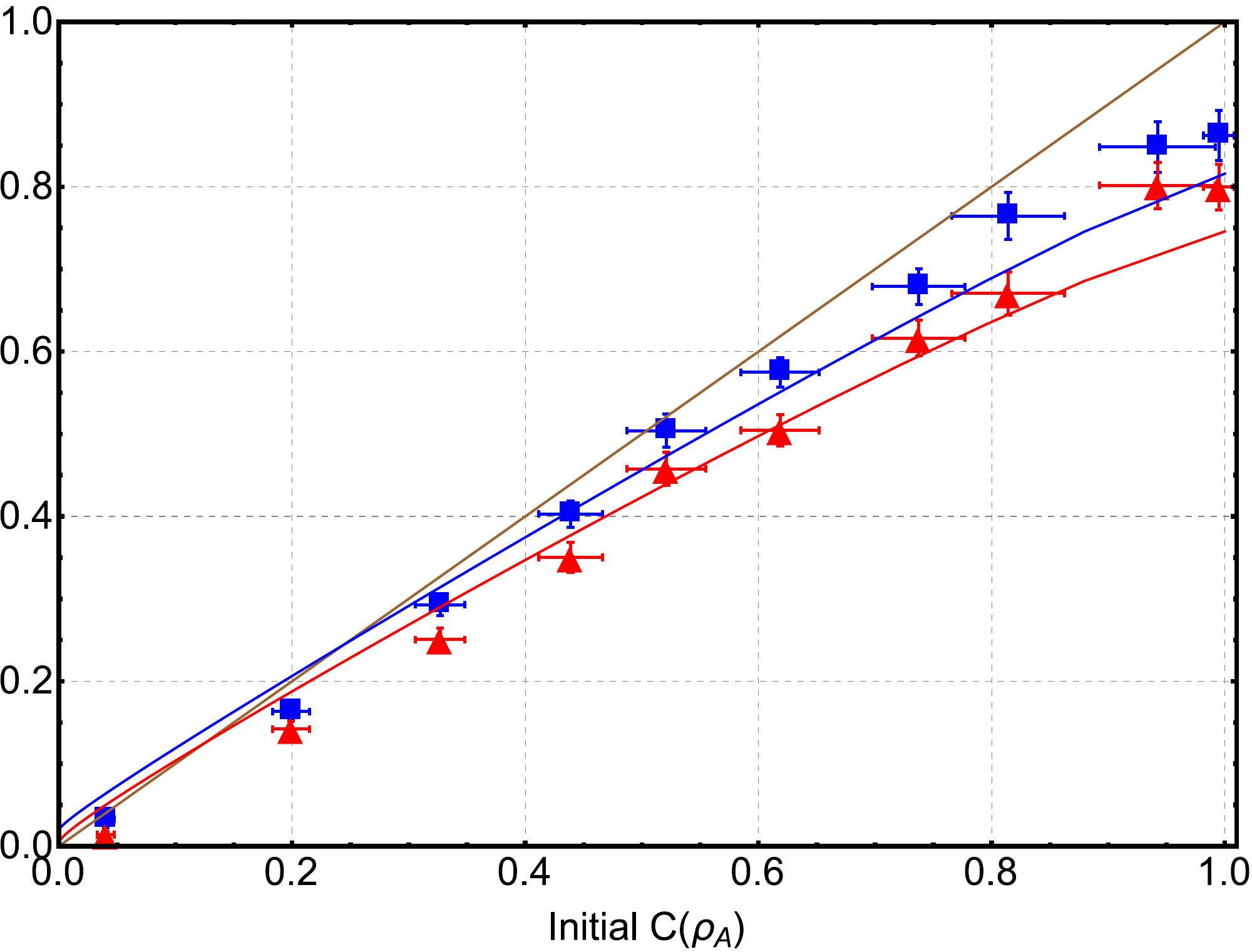}
\caption{\label{fig:mixedstate} \textbf{Discord generated and coherence restored for various mixed state inputs.} Input states of various purity are generated by subjecting a pure maximally coherent input $\ket{+}$ to quartz crystals of different thickness, resulting in mixed states $\rho_A$ with varying levels of initial coherence. The exact values are determined by quantum state tomography. For each crystal thickness, we plot the amount of discord successfully generate through the conversion process (blue squares) and the amount of coherence restored after one cycle (red triangle). Theoretical curves taking into account the imperfect fidelity of the CNOT gate (blue and red line) are plotted for comparison. Meanwhile the brown line represents the amount of coherence we would have achieved in the ideal case of perfect unitary gates (this of course, aligns with the initial coherence).}
\end{figure}

We also tested the inter-conversion of coherence and discord for mixed inputs of the form $\rho_A = \frac{1}{2}(I + a \ket{0}\bra{1} + a^{*} \ket{1}\bra{0})$ for various values of $a$. This involves significant modifications to the state preparation module, including the use of quartz crystals of different thickness to dephase an initially maximally coherent state. The resulting $\rho_A$ is then determined through quantum state tomography (see details in Supplementary Informations). The initial coherence can then be tuned from $0$ to $1$ by adjusting the thickness of quartz crystal (see Fig. \ref{fig:experiment}). Experimental results are illustrated in Fig. \ref{fig:mixedstate}. The amount of discord generated and the coherence restored after running one full cycle are depicted respectively by blue squares and red triangles. The qualitative behaviour matches that of the pure state scenario, where the amount of discord, and subsequent coherence are both degraded due to experimental imperfections. Numerical simulations taking the fidelity of our CNOT gate into account agrees well with experimental data (with details in Supplementary Informations).

\textbf{Discussion.} Here, we illustrated a cyclic scheme where coherence initially in a quantum system $A$ is consumed locally to synthesize an identical amount of discordant correlations with some ancilla $B$. These correlations are then harnessed to restore coherence in $A$. Under ideal conditions, this cycle is lossless, and can be repeated ad-infinitum. We realized one round of this cycle using linear optics, showing explicitly how coherence encoded within a photonic qubit can be converted to discord between it and an ancilla via incoherent operations. By measurement of the ancillary photon, we restored up to 80 percent of the coherence within the original qubit. Our experiment corroborates growing evidence non-classicality in correlations and non-classicality within singular quantum systems are closely connected.

%As pointed out in \cite{ConvertingCoherencetoQuantumCorrelations}, this conversion process is in fact taking place in some quantum algorithms. We expect our scheme would help us better understand the origin of the quantum advantage behind existing quantum algorithms, as well as develop novel quantum protocols exploiting quantum phenomena as the resource.

A full understanding of these connections is still in progress. While discord and coherence are popular quantifiers of non-classicality, there exists many other important variants. Ultimately, what constitutes a free state depends heavily on context. In continuous variable quantum optics, coherent states (not to be confused with the resource theory of coherence) are generally considered to be classical as they are easily synthesized -- despised being a superposition of energy eigenstates. Meanwhile in resource theories of quantum thermodynamics, thermal states are considered free, and anything else is a potential resource. Current research indicates these more diverse forms of non-classicality can also be inter-converted to non-classical correlations. For example, any state that's not a mixture of coherent states is a resource for generating entanglement using beamsplitters~\cite{wolf2003entangling,killoran2016converting}. Meanwhile conditional thermal operations were recently introduced to understand how measurement of suitably correlated ancilla can generate thermal resources~\cite{PhysRevA.95.012313}. %Indeed, unified treatments of these relations is exciting topic of research.

Another promising line of research is the role of such conversion processes in computation. Indeed, conversion of coherence to quantum correlations taking place in prominent quantum algorithms including both deterministic quantum computation with one quantum bit (DQC1) and factoring~\cite{ConvertingCoherencetoQuantumCorrelations}. Meanwhile the use of correlated resources, together with measurement feed-forward presents a strategy for implementing otherwise difficult to synthesize gates in measurement based computation~\cite{raussendorf2001one}. If the inter-conversion of various quantum resources could be related to certain operational properties of such protocols, it could suggests that operational benefits can emerge when one form of non-classicality is transformed into other.

\textbf{Acknowledgements} -- We thank V.Vedral, B. Yadin and X. Yuan for discussions. This work is supported by National Natural Science Foundation of China (Grant No. 11574291 and Grant No. 11622428), the National Key R \& D Program (Grant No. 2016YFA0301700 and Grant No. 2016YFA0301302), China Postdoctoral Science Foundation (Grant No. BH2030000024), the National Research Foundation of Singapore and in particular NRF Award No. NRF--NRFF2016--02 and CRP Award No. NRF-CRP14-2014-02, the John Templeton Foundation Grant 53914 {\em ``Occam's Quantum Mechanical Razor: Can Quantum theory admit the Simplest Understanding of Reality?''}, and the Foundational Questions Institute.

%\bibliographystyle{apsrev4-1}
%%\bibliographystyle{plain}
%
%\bibliography{all_references_hou}

%

\begin{appendix}

\section*{\label{sec:appendix1}S1. Theoretical Details}

Consider input state $\rho_{A}=\sum_{m,n=0}^{d_{A}-1}\rho_{A}^{mn}\ketbra{m}{n}$ on system $A$ with dimensions $d$. Choose an uncorrelated $d$ dimensional ancilla initialized in state $\tau_B$. Let $\rho^I_{AB} = \Delta_{IC}(\rho_{A}\bigotimes\tau_{B})$, where $\Delta_{IC}$ is some incoherent operation. From prior research, we that $D(\Delta_{IC}(\rho_{A}\bigotimes\tau_{B})) \leq C(\rho_{A})$. Furthermore, this inequality can always be saturated. To do so, we first transform $\tau_B$ to $\ketbra{0}{0}$ (this is always possible via incoherent operations), followed by application of a generalized controlled-NOT gate $\Delta_{CNOT}$, whereby $D(\rho^{Imax}_{AB}) = C(\rho_{A})$ and  $\rho^{Imax}_{AB} = \Delta_{CNOT}(\rho_{A} \bigotimes \ketbra{0}{0})$. Further details can be found in Ma et.al~\cite{ConvertingCoherencetoQuantumCorrelations}. This establishes correctness of Eq. (1).

\begin{figure*}[htp]\label{fig:QPT}
\includegraphics[scale=0.7]{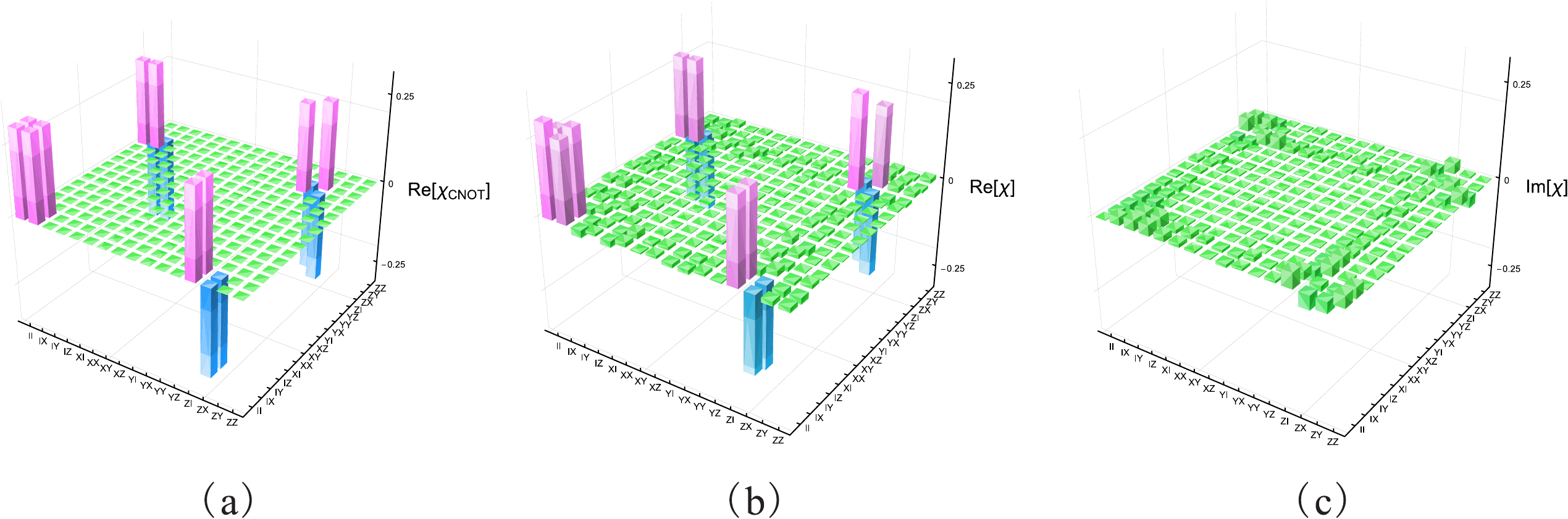}
\caption{\label{fig:qpt} \textbf{Characterization of experimentally realized CNOT gate.} We can fully specify the action of two-qubit gate by its process matrix elements $\tilde{\chi}_{mn}$, which defined it actions on a general input state $\rho$. Specifically, each $\rho$ is mapped to an output $\sum_{mn}\tilde{\chi}_{mn}\hat A_{m}\rho\hat A_{n}^{\dag}$, where the summation is over all possible two-qubit Pauli operators $A_k$. The process matrix for the ideal CNOT gate is plotted depicted in (a). Note that only the real elements are plotted as in the ideal case, all imaginary elements are zero. (b) and (c) respectively depict the real and imaginary elements of the process matrix for the CNOT gate realized in our experimental set-up. This yields a fidelity of 0.885.}
\end{figure*}

We now show for $\rho_{AB}^{Imax}$, the state which saturates Eq. (1), the coherence that can be restored to $A$ through LQICC operations is upper bounded by the quantum discord. That is
\begin{equation}\label{eqn:finalbound}
C(\rho_{A}^{\textmd{LQICC}})\leq D(\rho_{AB}^{Imax}) = C(\rho_A)\tag{S1},
\end{equation}
where $\rho_{A}^{\textmd{LQICC}}$ is the state of $A$ after LQICC operations. To do this, first recall a result in coherent distillation and steering~\cite{fanhuscireport2017,AssistedDistillationofQuantumCoherence}. Given an arbitrary bipartite quantum state $\rho_{AB}$, one has
\begin{equation}\label{cohbound2}
C(\rho_{A}^{\textmd{LQICC}})\leq C_{B|A}(\rho_{AB})\tag{S2},
\end{equation}
where $\rho_{A}^{\textmd{LQICC}}$ is the state of $A$ after some arbitrary LQICC operation, $C_{B|A}(\rho_{AB})=\min_{\chi_{AB}\in\mathcal{QI}}S(\rho_{AB}||\chi_{AB})=S(\Phi_{A}^{i}(\rho_{AB}))-S(\rho_{AB})$ is the quantum-incoherent (QI) relative entropy of $\rho_{AB}$. Here $\mathcal{QI}$ denotes the set of quantum incoherent states which take the form $\chi_{AB}=\sum_{i}p_{i}\ketbra{i}{i}_{A}\otimes\sigma^{B}_{i}$, and $\Phi^{i}_{A}$ represents the dephasing operator in the reference basis $\{\ket{i}_{A}\}$. The second component we need is the following lemma:

\begin{lemma} The QI relative entropy created between a state $\rho_A$ and an incoherent ancilla $\tau_B$ by an incoherent operation $\Delta_{IC}$ is upper bounded by the coherence of $\rho_A$:
\begin{equation}\label{cohbound2}
C_{B|A}(\Delta_{IC}(\rho_A\otimes\tau_B))\leq C(\rho_A)\tag{S3},
\end{equation}
When the dimensions of $A$ and $B$ coincide, this bound can be saturated by setting $\Delta_{IC} = \Delta_{CNOT}$, .
\begin{proof} Suppose $\tau_A$ is the closest incoherent state to $\rho_A$. By exploiting the contractivity of quantum relative entropy and that $\Delta_{IC}(\tau_A\otimes\tau_B)\in\mathcal{QI}$, we have $C(\rho_A)=S(\rho_A || \tau_A)=S(\rho_A\otimes\tau_B||\tau_A\otimes\tau_B)\geq S(\Delta_{IC}(\rho_A\otimes\tau_B)||\Delta_{IC}(\tau_A\otimes\tau_B))\geq C_{B|A}(\Delta_{IC}(\rho_A\otimes\tau_B))$. When $A$ and $B$ have the same dimension, this bound can be saturated by a generalized CNOT operation. Specifically, let $\tau_B=\ketbra{i_0}{i_0}$. There then exists a generalized CNOT operation $\Delta_{CNOT}$ such that $E(\Delta_{CNOT}(\rho_A\otimes\tau_B))=C(\rho_A)$, where $E$ is the relative entropy of entanglement~\cite{MeasuringQuantumCoherencewithEntanglement}. Since $C_{B|A}(\rho)\geq E(\rho)$, one has $C_{B|A}(\Delta_{IC}(\rho_A\otimes\tau_B))\geq C(\rho_A)$. Thus $C_{B|A}(\Delta_{CNOT}(\rho_A\otimes\tau_B))= C(\rho_A)$. If $\tau_B$ is any other incoherent state, one first maps it to $\ketbra{i_0}{i_0}$, which is an incoherent operation, and the proof follows.
\end{proof}
\end{lemma}

From Lemma 1 and $D(\rho_{AB}^{Imax}=\Delta_{CNOT}(\rho_A\otimes\tau_B)) \leq C(\rho_A)$, one has $C_{B|A}(\rho_{AB}^{Imax})=D(\rho_{AB}^{Imax})$. Combining this with Eq. \eqref{cohbound2}, one has $C_{B|A}(\Delta_{IC}(\rho_A\otimes\tau_B))\leq D(\rho_{AB}^{Imax})$, for any incoherent operation $\Delta_{IC}$. Therefore we have $C(\rho_{A})=C_{B|A}(\rho_{A}\otimes\tau_{B}) = D(\rho_{AB}^{Imax})\geq C_{B|A}(\rho_{AB}^{I})\geq C(\rho_{A}^{\textmd{LQICC}})$, which establishes that the coherence in $A$ at the end of one cycle cannot exceed the amount of coherence in $A$ at the beginning of the cycle, implying Eq \eqref{eqn:finalbound}.

This proves that this bound can always be saturated, consider the case where we adopt the optimal strategy in the conversion phase (i.e., through application of $\Delta_{CNOT}$) such that we obtain $\rho^{Imax}_{AB}$. We now establish that there exists a measurement, specified by a projectors $\ketbra{\Psi_{j}}{\Psi_{j}}$ that can be performed on $B$, together with possible classical communications and local incoherent operations on $A$ that allow us to steer $A$ into a state with coherence $C(\rho_{A})$. First observe that $\rho_{AB}^{Imax}$ has the form of maximally correlated state $\sum_{m,n}\rho_{A}^{mn}\ketbra{mm}{nn}$. Now let $\ket{\Psi_{j}}= \sum_{k=0}^{d-1}e^{-2 \pi i j k/d}\ket{k}$. Each measurement outcome $j$ occurs with equiprobability, collapsing $A$ to the corresponding state $\sum_{m,n=0}^{d-1}\rho_{mn}^{A}e^{2 \pi ij(m-n)/d}\ketbra{m}{n}$, with coherence of exactly $C(\rho_{A})$. To see this, note that performing the incoherent operation $\Delta_i = \sum_k e^{-2\pi i k}\ketbra{k}{k}$  conditioned on the measurement outcome $i$ will steer $A$ back to the original input state $\rho_A$. i.e., $\rho_{A}^{\textmd{LQICC}}=\rho_{A}$.

\section*{\label{sec:appendix2}S2. Experimental Details}

As illustrated in Fig. 2, the experimental setup is divided into three modules (a) state preparation, (b) coherence to
discord conversion and (c) coherence restoration. In the state preparation module, a source polarization entangled photon pairs is created using 390 nm femto-second pump light, frequency-doubled from a 780 nm mode-locked Ti:sapphire pulsed laser with a pulse width of approximately 150 fs and repetition rate 76 MHz. The laser is used in the down-conversion process using sandwich like type-II beamlike phase-matching ¦Â-barium borate (BBO) crystals~\cite{beamlikeZhangchao}. Desired pure product states $\ket{\phi}_{A}\ket{0}_{B}$ can then be created using a pair of polarizing beamsplitters, and a suitably positioned half-wave plate (HWP1). To prepare the more general mixed states used in the protocol, we make use of quartz crystals to decohere an initially maximally coherent state on system $A$. Note that a separate quartz crystals with the same length is inserted in path $B$ to compensate for the resulting change in optical path for best interference in subsequent stages of the experiment.

In the coherence to discord conversion module, we implement an all optical CNOT gate (see Fig. 2 b)~\cite{LinearOpticsControlledPhaseGateMadeSimple,DemonstrationofanOpticalQuantumControlledNOTGatewithoutPathInterference}. This is constructed by three partially polarizing beam splitters (PPBS) and six HWPs, and has a success rate of $1/9$. The thickness of aforementioned quartz crystals are tuned for interference between the two incoming photons. The final coherence restoration module involves the use of two single photon detectors, a polarizing beam-splitter and suitably tuned wave plates to perform local polarization measurements on $B$.

Experimental results at each stage are recovered by quantum state tomography. For detecting initial coherence on $A$, the state of $A$ is analyzed directly by single-qubit quantum state tomography. Thus allows us to estimate the coherence of each mixed input after dephasing by aforementioned quartz crystals directly from tomographic data. At end of the conversion module (see Fig. 2 b), we perform full tomography to reconstruct the density matrix $\rho_{AB}^{Imax}$, which is then used to measure the resulting discord generated through the conversion process. Similarly tomography, tomography was performed after the coherence restoration module on system $A$ to characterize how much coherence was restored after one inter-conversion cycle. For characterization of imperfections in our CNOT gate, we estimate its process matrix using the maximum-likelihood method~\cite{Jeif03quantum}. The results are indicated in Figure. \ref{fig:qpt}, where the fidelity of the gate is estimated to be $0.885$. Taking this error into account allows us to match theory with experimentally observed results.

\section*{S3. Experimental Data}

\begin{table}[htp!]
\caption{Experimental Data for pure state}
\begin{tabular}{c|ccccc}
\hline
\hline
$\theta$& 2& 5& 8& 11& 14\\
\hline
$D$&0.121& 0.248& 0.4& 0.543& 0.7\\
\hline
$C_F$&0.114& 0.232& 0.394& 0.526& 0.63\\
\hline
\hline
$\theta$& 17& 20& 23& 26& 29\\
\hline
$D$& 0.767& 0.803& 0.823& 0.758& 0.689\\
\hline
$C_F$& 0.751& 0.784& 0.773& 0.671& 0.649\\
\hline
\hline
$\theta$& 32& 35& 38& 41& 44\\
\hline
$D$& 0.572& 0.462& 0.312& 0.157& 0.042\\
\hline
$C_F$& 0.505& 0.429& 0.305& 0.127& 0.042\\
\hline
\hline
\end{tabular}
\end{table}

\begin{table}[htp!]
\caption{Experimental Data for mixed state}
\begin{tabular}{c|ccccc}
\hline
\hline
$l$&200 &120 &100 &90 &76\\
\hline
$C_I$&0.04& 0.199& 0.327& 0.439& 0.521\\
\hline
$D$&0.032& 0.163& 0.293& 0.403& 0.504\\
\hline
$C_F$&0.014& 0.142&0.251&0.35&0.458\\
\hline
\hline
$l$&60 &46 &30 &16 &0\\
\hline
$C_I$& 0.618& 0.737& 0.815& 0.943& 0.995\\
\hline
$D$& 0.575& 0.679& 0.764& 0.848& 0.862\\
\hline
$C_F$&0.504&0.616&0.671&0.802&0.8\\
\hline
\hline
\end{tabular}
\end{table}

Here we list the experimental data collected. Initial coherence and discord after application of CNOT gate are retrieved trough aforementioned tomographic procedure. The final coherence in $A$ involves a minor addendum as in general, the coherence obtained depends of the result of measuring $B$. Here, our final coherence is estimate by taking a weighted average, such that $C_F = p_+ C^+_F +  p_- C^-_F$, where $p_{+}$ and $p_-$ respectively denote the proportion coincidence counts in which $B$ was collapsed into $\ket{\pm}_y$.

Here, $D$ represents the discord that successfully converted from coherence and $C_F$ the final coherence after one inter-conversion cycle. Fore pure states, these quantities are tabulated against $\theta$, which represents experiments where the initial resource state is given by $\ket{\phi} = \cos(2\theta)\ket{0} + \sin(2\theta)\ket{1}$. For experiments with mixed states, these quantities a tabulated against $l$, represents the thickness of the decohering quantz crystal (in terms of number of wavelengths). In addition, we include $C_I$, representing the initial coherence of the photon $A$, as determined by tomography.
\end{appendix}

%\textbf{Author contributions}

\end{document}